%% file: paper.tex
\documentclass{easychair}
\usepackage{multirow}
\title{Revisiting Decision Diagrams for SAT%
\thanks{Supported by Austrian Science Fund (FWF), NFN S11408-N23 (RiSE), and the Institutional Strategy of the University of Bremen, funded by the German Excellence Initiative.}}
\author{Tom van Dijk\inst{1} \and R\"udiger Ehlers\inst{2} \and Armin Biere\inst{1}}
\institute{%
Johannes Kepler University Linz, Austria \\
\email{tom.vandijk@jku.at}~~\email{biere@jku.at}
\and
University of Bremen, Germany \\
\email{rehlers@uni-bremen.de}}
\authorrunning{Van Dijk, Ehlers and Biere}
\titlerunning{Revisiting Decision Diagrams for SAT}

\usepackage[latin1]{inputenc}

\usepackage{amsopn}
\usepackage{amsmath}
\usepackage{amsfonts}
\usepackage{graphicx}
\usepackage{pgf}
\usepackage{tikz}
\usepackage{url}
\usetikzlibrary{arrows,automata,shapes}
\DeclareMathOperator{\subdiff}{\tikz[baseline=0.1em]{\path [draw] (0,0) -- (0.1,0) -- (-0.15,0.25) -- (-0.25,0.25) -- (0,0);}}

\newcommand{\newterm}[1]{\textit{#1}}

\begin{document}
\maketitle
\begin{abstract}
Symbolic variants of clause distribution using decision diagrams to
eliminate variables in SAT were shown to perform well on hard combinatorial
instances.  In this paper we revisit both existing ZDD and BDD variants of
this approach.  We further investigate different heuristics for selecting
the next variable to eliminate.  Our implementation makes further use of
parallel features of the open source BDD library Sylvan.
\end{abstract}

\section{Introduction}

In the past, decision diagrams have been used for the Satisfiability (SAT) problem in different ways.
To compute all satisfiable models for a set of clauses one can build the binary decision diagram (BDD) of the conjunction of all clauses.
In general, this is expensive and only works for very small or specific models.

Most approaches to SAT manipulate the set of clauses until a satisfying solution is found or unsatisfiability is established by deriving the empty clause.
Decision diagrams have also been used in the past to represent the set of clauses,
in particular by the \textsc{ZRES} solver by Chatalic and
Simon~\cite{DBLP:journals/ijait/ChatalicS01}.
For a more complete list of references see also~\cite{DBLP:conf/csr/SinzB06}.

Chatalic and Simon were able to show that SAT instances encoding the problem
of putting $n + 1$ pigeons into $n$ holes, proven
to be hard for any resolution-based SAT solver~\cite{DBLP:journals/tcs/Haken85}, can be solved in polynomial time using a compressed \newterm{conjunctive normal form} (CNF) clause representation with \newterm{zero-suppressed binary decision diagrams} (ZDDs) \cite{DBLP:conf/dac/Minato93}.
They defined advanced clause set disjunction and conjunction operators on ZDDs that remove subsumed clauses during the variable elimination process.
This way, the number of diagram nodes grows only polynomially during solving.

Sets of clauses are concisely represented using ZDDs as the number of ZDD nodes required is strictly smaller than or equal to the total number of literals in the clauses.
Because decision diagrams share isomorphic subgraphs, ZDD nodes can be reused.
Under certain conditions the ZDD representing a set of clauses can be significantly smaller, essentially compressing the set of clauses by a factor of $\#{nodes}$/$\#{literals}$.
We call this ratio the \newterm{compression ratio}.

The relatively little interest in ZDD-based clause representations contrasts with their high usability as a basis for experimenting with advanced simplification techniques for clause sets.
In this paper, we revisit these methods and study several new techniques.

\section{Background}

Chatalic and Simon \cite{DBLP:journals/ijait/ChatalicS01} employ ZDDs whose
variables are the literals in the original set of clauses.
Hence, the number of variables in the ZDD is twice as large as in the SAT instance that the ZDD encodes. A clause is contained in the clause set represented by a ZDD if there exists a path from the ZDDs root to the $1$ sink node along which exactly the ZDD variables that represent the literals in the clause have $\mathbf{true}$ values.

Chatalic and Simon show how to perform variable elimination on such clause sets, which allows to solve the SAT problem by just eliminating all variables. To speed up the solution, they also define a \emph{subsumption removal} operation, which removes all clauses from a clause set $S$ represented in ZDD form that are subsumed by other clauses in $S$.
This technique is a representative for \newterm{clause level techniques}, which aim at reducing or simplifying the set of clauses. Subsumption removal does not necessarily reduce the number of nodes in a ZDD and hence does not guarantee improvements in the further ZDD operations. In practice, non-improving removals are however rare. 

Inspired by the success of bounded variable elimination for CNF~\cite{DBLP:conf/sat/EenB05}, we also looked into bounding the nodes in ZDD-based variable elimination or alternatively the size of the CNF represented by a ZDD.
The idea is to limit the growth of the ZDD during variable elimination.
If speculatively eliminating a variable increases the size too much, its elimination is undone.

Finally, in recent work, parallel implementations of BDDs have been proposed, in particular the package Sylvan~\cite{DBLP:conf/tacas/DijkP15,DBLP:phd/basesearch/Dijk16} which parallelizes the low-level BDD operations. We look at how this affects the approach of building the BDD of satisfiable models based from the CNF.

\section{Solving SAT by Computing the BDD}

The naive approach to solving SAT symbolically encodes each clause as a BDD and then computes the conjunction of all encoded clauses.
We use a slightly different approach, by first encoding the CNF as a ZDD and then computing the BDD of the models from the ZDD.
In this way, BDDs of shared ZDD nodes can be cached and only need to be computed once.
There is furthermore a potential to skip redundant BDD nodes.
Our implementation is based on the parallel decision diagram package Sylvan~\cite{DBLP:conf/tacas/DijkP15,DBLP:phd/basesearch/Dijk16} which uses work-stealing to implement parallel operations.

Sylvan parallelizes BDD operations for multi-core systems using two techniques.
The first technique is to use a work-stealing framework to perform load-balancing on multiple cores.
As BDD operations are implemented as recursive tasks that typically spawn two independent subtasks, such operations naturally form a tree of tasks that can be executed using techniques like work-stealing.
The reason Sylvan uses work-stealing is because work-stealing has been demonstrated to be very efficient and scalable for fine-grained task parallelism.
BDD operations mainly consist of calls to two hash tables: the operation cache and the unique nodes table.
The second technique is to use specialized lock-free hash tables for the operation cache and the unique nodes table to maximize the scalability of the BDD operations.

In our implementation of converting a CNF to a BDD, we first
read all clauses to an array of integers, and then we recursively
encode this CNF to a ZDD in a straight-forward manner.
We then use another algorithm that computes an equivalent BDD for each ZDD node. The BDD for the ZDDs root node then represents the BDD for the set of solutions to the SAT problem.
Both the encoding of the CNF to a ZDD and computing the BDD from this ZDD are performed in parallel,
by using our parallel decision diagram
library Sylvan~\cite{DBLP:conf/tacas/DijkP15,DBLP:phd/basesearch/Dijk16}.

\begin{table}[t]
\input{experiments/bdds/direct/table.tex}
\\[2ex]
\input{experiments/bdds/indirect/table.tex}
\caption{%
In the upper table we compute the BDD directly during parsing of the DIMACS file,
while in the bottom table we read the file into a ZDD and then compute 
a BDD recursively.
}
\label{table:bdds}
\end{table}

The experiments in Table~\ref{table:bdds}
were performed using 32-core Xeon E5-2620 machines, with 128 GB available memory,
although we only allocated 2 GB for the experiments.
The experiments show reasonable speed-ups up to 19 pigeons.
For 20 pigeons we get strange results, which still need to be
investigated.
We may also want to look at other examples in the future.

\section{Operations on ZDDs for Clause Sets}

In this section, we review the operations on ZDDs described by Chatalic and Simon \cite{DBLP:journals/ijait/ChatalicS01}. We also introduce and apply novel simplifications for the presentation of operations on ZDDs that represent sets of clauses. For example, for the \newterm{clause intersection} operation which corresponds to logical disjunction on clause sets, six of the seven non-final cases of the recursion procedure can be dropped. 

A ZDD representing a clause set is a directed acyclic graph (DAG) with a $0$ sink, a $1$ sink, and a designated root node (which can be one of the two sinks as special cases). All nodes except for the two sinks are labelled by either $v$ or $\neg v$ for some set of SAT variables $\mathcal{V}$. There exists an ordering $O = [v_1, \neg v_1, \ldots, v_n, \neg v_n]$ of the possible node labels, and along every path from the root to a sink node, node labels can occur only once and in the order defined by $O$. Not every node label has to occur along every path, though. Every node except for the sinks has exactly two successors, the $\mathit{then}$- and $\mathit{else}$-successors. The ZDD represents exactly the clauses represented by the paths from the root to the $1$ sink, where whenever a $\mathit{then}$-edge is taken from a node labeled by $l$ for $l \in \mathcal{V} \cup \{\neg v \mid v \in \mathcal{V}\}$, then $l$ is a literal in the clause. We also view an order $O$ as a function that maps a literal to its position in the order.

A ZDD is said to be in \newterm{reduced form} if (1) for no two nodes $n_1$ and $n_2$ in the DAG with the same label, the $\mathit{then}$-successor of $n_1$ is the same as the $\mathit{then}$-successor of $n_2$ and the $\mathit{else}$-successor of $n_1$ is the same as the $\mathit{else}$-successor of $n_2$, and (2) the $\mathit{then}$-successor of no node is the $0$ sink. We also call these requirements the \newterm{reduction rules of ZDDs} and without loss of generality, we will consider reduced ZDDs in the following, as practical implementations of ZDDs and their operations normally operate on these.

Given a $v$-labelled node in the graph with a $\mathit{then}$-successor $n_1$ and the $\mathit{else}$-successor $n_2$, we denote the node using the notation $\bigtriangleup(v,n_1,n_2)$. We consider only ZDDs in which for no variable $v$, both $v$ and $\neg v$ can occur as literals in a clause. Such clauses are redundant and if we ensure that all operations we apply to ZDDs in the following never introduce such clauses, this assumption is justified. We make use of it because it allows us to simplify notation.
In many cases, a $v$-node has a $\neg v$-node successor for some $v \in \mathcal{V}$. In this case, we introduce \[\bigtriangledown(v,n_1,n_2,n_3) = \bigtriangleup(v,n_1,\bigtriangleup(\neg v,n_2,n_3))\] as a more elegant description of this node pair. Additionally, as a node of the form $\bigtriangleup(\neg v,n_1, n_2)$ is semantically equivalent to $\bigtriangleup(v,0, \bigtriangleup(\neg v,n_1, n_2))$, we define \[\bigtriangledown(v,0,n_1,n_2) = \bigtriangleup(v,0, \bigtriangleup(\neg v,n_1, n_2)) = \bigtriangleup(\neg v,n_1, n_2)\] for $v \in \mathcal{V}$ and $\bigtriangledown(v,n_1,0,n_2) = \bigtriangleup( v,n_1, n_2)$ if $\tau(n_2) \neq \neg v$, where $\tau$ maps every node to its label.

By using these definitions, the application of all operations on ZDD nodes only needs to be described for \newterm{combined nodes} of the form $\bigtriangledown(v,n_1,n_2,n_3)$ as the cases that a $v$-node has no $\neg v$ successor and a $\neg v$ node has no $v$ predecessor are now special cases and need not be explicitly considered for all functions respecting the clause-set semantics of ZDDs.
In the following, for every combined node $n = \bigtriangledown(v,n_1,n_2,n_3)$, we call $n_1$ the then-successor of $n$, $n_2$ the else-successor of $n$ and $n_3$ the don't care-successor of $n$. We furthermore use sets of clauses and their respective ZDDs interchangeably.

ZDD operations are -- as BDD operations -- normally defined as recursive functions operating on two decision diagrams. They recurse on pairs of ZDD nodes (one for each input ZDD), where the two nodes can have different labels.
For the binary operations to follow, we only need to state the result of their applications to pairs of nodes with the same labels as we can always introduce additional (artificial helper) ZDD nodes while not changing any clause, i.\,e.\ for any operation $\odot$ respecting the semantics of the ZDDs, we have \[\bigtriangledown(v,n_1,n_2,n_3) \odot \bigtriangledown(v',m_1,m_2,m_3) = \bigtriangledown(v,n_1,n_2,n_3) \odot \bigtriangledown(v,0,0,\bigtriangledown(v',m_1,m_2,m_3))\] if $O(v)<O(v')$ and   \[\bigtriangledown(v,n_1,n_2,n_3) \odot \bigtriangledown(v',m_1,m_2,m_3) = \bigtriangledown(v',0,0,\bigtriangledown(v,n_1,n_2,n_3)) \odot \bigtriangledown(v',m_1,m_2,m_3)\] if $O(v)>O(v')$. 

For the actual implementation of functions on ZDDs, it however makes sense to consider these special cases in order to speed up the computation. Additionally, implementations of these functions must apply the ZDD reduction rules after the computation of $\bigtriangledown(v,n_1,n_2,n_3) \odot \bigtriangledown(v',m_1,m_2,m_3)$ in order to obtain a valid ZDD. 

As a first example for a ZDD operation, a \newterm{clause union operator} $\sqcup$, which merges two sets of clauses in ZDD form to a resulting ZDD containing the clauses of both sets, can be described (for $X$ being a place holder for either an internal node or a sink) as follows (for all $v \in \mathcal{V}$):
\begin{align*}
0 \sqcup X &= X \\
X \sqcup 0 &= X \\
1 \sqcup X &= 1 \\
X \sqcup 1 &= 1 \\
\bigtriangledown(v,n_1,n_2,n_3) \sqcup \bigtriangledown(v,m_1,m_2,m_3) &= \bigtriangledown(v,n_1 \sqcup m_1,n_2 \sqcup m_2 ,n_3 \sqcup m_3)
\end{align*}
In this definition, the first two basic cases reflect the case that during the recursive application of this procedure, there is no clause in the one set corresponding to the current clause prefix(es) in the other set. In this case, the respective part from the other set can simple be copied. The third and fourth case represent situations in which for one ZDD, all literals of a clause have already been seen along a path in the ZDD while in the second ZDD, there exists a clause containing the same but possibly additional literals. This might for example occur if we merge the sets of clauses $\{a \vee b \vee c\}$ and $\{a \vee b \vee c \vee d\}$. In such a case we can safely drop the additional literals as the longer clause would be subsumed by the shorter one. The final case involving the inner nodes simply splits the set of clauses to be represented by the occurrence of a positive or negative literal for the current variable in the variable order.

Chatalic and Simon \cite{DBLP:journals/ijait/ChatalicS01} also defined 
a subsumption removal operator that can be defined (for $X$ being a place holder for either an inner node or a sink) as follows (for all $v \in \mathcal{V}$):
\begin{align*}
0 \subdiff X &{\quad=\quad} 0 \\
1 \subdiff 1 &{\quad=\quad} 0 \\
1 \subdiff \bigtriangledown(v,n_1,n_2,n_3) &{\quad=\quad} 1 \\
\bigtriangledown(v,n_1,n_2,n_3) \subdiff 1 &{\quad=\quad} 0 \\
\bigtriangledown(v,n_1,n_2,n_3) \subdiff 0 &{\quad=\quad} \bigtriangledown(v,n_1,n_2,n_3) \\
\bigtriangledown(v,n_1,n_2,n_3) \subdiff \bigtriangledown(v,m_1,m_2,m_3) &{\quad=\quad} \bigtriangledown(v,(n_1 \subdiff m_1) \subdiff m_3, \\
& \phantom{\quad=\quad} (n_2 \subdiff m_2) \subdiff m_3 ,n_3 \subdiff m_3) 
\end{align*}
By taking $A' = A \subdiff B$ for some clause sets $A$ and $B$, a modified set of clauses $A'$ is computed that is equivalent to $A$, with the only difference that all clauses that are subsumed by some other clause in $B$ have been removed. A clause $l_1 \vee \ldots \vee l_m$ is said to be subsumed by a clause $l'_1 \vee \ldots \vee l'_{m'}$ if $\{l_1,\ldots,l_m\} \supseteq \{l'_1,\ldots,l'_{m'}\}$. The $\subdiff$ operator can be used to define a subsumption removal function $\mathit{SF}$ that removes all clauses from a clause set $A$ that are subsumed by other clauses in $A$, making sure that $\mathit{SF}(A)$ is subsumption-free:
\begin{align*}
\mathit{SF}(0) &{\quad=\quad} 0 \\
\mathit{SF}(1) &{\quad=\quad} 1 \\
\mathit{SF}(\bigtriangledown(v,n_1,n_2,n_3)) & {\quad=\quad} \bigtriangledown(v,\mathit{SF}(n_1) \subdiff \mathit{SF}(n_3) ,\mathit{SF}(n_2) \subdiff \mathit{SF}(n_3) ,\mathit{SF}(n_3))
\end{align*}
The following clause union function combines two sets of clauses in ZDD form to a single set containing the clauses of both of the operand sets (corresponding to logical conjunction). We only state a version of this operation that makes sure that subsumed clauses are removed if the input sets are subsumption-free. In particular, for $A \sqcup_S B$, all clauses in $A$/$B$ that are subsumed by some other clause in $B$/$A$ are removed, respectively. We use the term \newterm{cross-subsumption} for such cases.
\begin{align*}
0 \sqcup_S X &{\quad=\quad} X \\
1 \sqcup_S X &{\quad=\quad} 1 \\
X \sqcup_S 0 &{\quad=\quad} X \\
X \sqcup_S 1 &{\quad=\quad} 1 \\
\bigtriangledown(v,n_1,n_2,n_3) \sqcup_S \bigtriangledown(v',m_1,m_2,m_3) &{\quad=\quad} \bigtriangledown(v,(n_1 \sqcup_S m_1) \subdiff (n_3 \sqcup_S m_3), \\
& \phantom{\quad=\quad} (n_2 \sqcup_S m_2) \subdiff (n_3 \sqcup_S m_3), n_3 \sqcup_S m_3)  \\
\end{align*}
Finally, an operation corresponding to logical disjunction is to be defined. Below is a definition of a suitable \newterm{clause distribution} operator which also makes sure that cross-subsumed clauses are removed.
\begin{align*}
0 \times_S X &{\quad=\quad} 0 \\
1 \times_S X &{\quad=\quad} X \\
X \times_S 0 &{\quad=\quad} 0 \\
X \times_S 1 &{\quad=\quad} X \\
& \hspace{-1cm}\bigtriangledown(v,n_1,n_2,n_3) \times_S \bigtriangledown(v',m_1,m_2,m_3) \quad = \\
& \bigtriangledown(v,((n_1 \times_S m_1) \sqcup_S (n_1 \times_S m_3) \sqcup_S (n_3 \times_S m_1)) \subdiff (n_3 \times_S m_3), \\
& ((n_2 \times_S m_2) \sqcup_S (n_2 \times_S m_3) \sqcup_S (n_3 \times_S m_2)) \subdiff (n_3 \times_S m_3), 
(n_3 \times_S m_3))
\end{align*}

\section{Variable elimination}

In their work on ZDDs for SAT solving, Chatalic and Simon~\cite{DBLP:journals/ijait/ChatalicS01} implemented the Davis-Putnam (DP) method, i.e., solving SAT by variable elimination, which works particularly well for the pigeon
hole problem.
In the DP procedure, variables are eliminated one by one in some order.
The heuristic for choosing the variable to be eliminated is not described precisely in
the work by Chatalic and Simon, except that the authors claim that their implementation tends to maximize the number of clauses (represented by the ZDD) produced at each elimination step.

We implemented our variant of their approach as follows.
Given a variable $v$ to eliminate, we first \emph{extract} a ZDD $A_v$ representing all
clauses which involve $v$ both positively and negatively.
The ZDD of $A_v$ is then split into a ZDD $A^{+}_v$ representing clauses with positive
occurrence and another ZDD $A^{-}_v$ representing clauses with negative
occurrence (removing $v$ and $\neg v$).
Then we apply the \emph{clause distribution} operation described above to obtain
a ZDD representing all the resolvents on the chosen variable, i.e.,
$A^{+}_{v} \times_S A^{-}_{v}$.
Next, the ZDD $A_v$ is subtracted from the ZDD for the whole CNF.
Finally the resulting ZDD from clause distribution is added through
the \emph{clause union} operation also described above.
Note that during both clause distribution and the union operation, subsumed clauses are
removed automatically and all operations are recursively parallelized.

We use three strategies to select the order of variables to eliminate next.  In
our \emph{original} variant, we simply picked variables in the order they
are listed in the DIMACS input file that contains the SAT instance (as integers).
In the second and third
approaches we perform the procedure described above speculatively and only
keep the result if the result stays within a certain \emph{bound}.
As bound we either use the number of \emph{nodes} of the ZDD (in the second
approach) or the number of represented \emph{clauses} of the ZDD (in the
third approach).

If speculative eliminations fails for \emph{all} variables, then our procedure
simply eliminates the first variable with the lowest increase (in either
nodes or clauses).  We continue with the original bound afterwards, which in
our experiments was always kept at zero.  Thus, eliminations that reduce the
number of nodes or clauses are always performed eagerly, while more costly
eliminations are delayed.

\begin{table}
\input{experiments/zdds/table.tex}
\caption{ZDD-based variable elimination.}
\label{table:zbdds}
\end{table}

The results are shown in Table~\ref{table:zbdds} for all three strategies.
Clearly, bounded variable elimination minimizing the number of nodes works best.  It scales up
to 20 pigeons.  However, it is slower than simple BDD-based conjunction~(cf.~Table~\ref{table:bdds}).
Also note that parallel speed-ups are only observed for the otherwise
inferior strategies (``original'' and ``clause'').

\section{Conclusion}

We revisited BDD-based solving and ZDD-based variable elimination for
hard combinatorial benchmarks. We also provide a reimplementation
in a parallel BDD library.  Our experiments on pigeon hole formulas show reasonable
speed-ups on multi-core machines.
Beside aspects related to parallelism we investigated two natural bounding
schemes to select the elimination order of variables.

We have also applied Quine-McCluskey style resolution and computation of irredundant
clause sets in this setting.
However, our preliminary results on
optimizing the size of ZDDs in this way and translating them back into CNF only
produced inconclusive experimental results, and thus are not reported in this
paper.

It is unclear how to generate proofs for these procedures, with the goal to
increase the level of confidence in these results.
We want to further study complexity and compression ratio of these procedures, both
theoretically on specific problems, as well as empirically on more
problem instances.
On the practical side, it might be interesting to also parallelize
searching for the best variable to eliminate.

\bibliographystyle{plain}
\bibliography{paper}

\end{document}

%% file: experiments/bdds/direct/table.tex
\begin{tabular}{|rrrrrrrrrrrr|}
\hline
cores
& ph10& ph11& ph12& ph13& ph14& ph15& ph16& ph17& ph18& ph19& ph20\\
\hline
1& 0.43& 0.46& 0.50& 0.61& 0.90& 1.54& 3.02& 6.31& 18.18& 37.15& 58.95\\
2& 0.62& 0.63& 0.73& 0.82& 1.04& 1.19& 2.20& 4.29& 9.61& 22.27& 33.38\\
4& 0.61& 0.63& 0.67& 0.72& 0.87& 1.17& 1.77& 3.15& 8.45& 18.55& 143.05\\
6& 0.37& 0.41& 0.39& 0.47& 0.59& 0.83& 1.31& 2.44& 5.87& 12.50& 98.28\\
8& 0.70& 0.71& 0.76& 0.81& 0.87& 1.06& 1.45& 2.45& 5.04& 10.98& 132.32\\
10& 0.79& 0.84& 0.87& 0.94& 1.06& 1.30& 1.98& 2.88& 5.63& 10.52& 93.82\\
12& 0.26& 0.29& 0.32& 0.42& 0.54& 0.83& 1.28& 2.35& 5.27& 10.06& 108.49\\
14& 0.34& 0.37& 0.42& 0.48& 0.63& 0.86& 1.34& 2.31& 4.83& 9.22& 102.03\\
16& 0.81& 0.81& 0.86& 0.91& 1.05& 1.31& 1.79& 2.72& 4.92& 9.01& 57.23\\
\hline
\end{tabular}

%% file: experiments/bdds/indirect/table.tex
\begin{tabular}{|rrrrrrrrrrrr|}
\hline
cores
& ph10& ph11& ph12& ph13& ph14& ph15& ph16& ph17& ph18& ph19& ph20\\
\hline
1& 0.93& 0.96& 1.00& 1.11& 1.39& 2.04& 3.52& 6.80& 18.80& 38.46& 54.50\\
2& 0.94& 0.94& 1.00& 1.08& 1.27& 1.67& 2.53& 4.62& 11.27& 23.04& 37.13\\
4& 0.94& 0.96& 0.98& 1.04& 1.17& 1.47& 2.02& 3.61& 8.99& 16.36& 29.42\\
6& 0.47& 0.47& 0.48& 0.53& 0.63& 0.95& 1.37& 2.44& 6.96& 14.07& 105.88\\
8& 0.18& 0.17& 0.20& 0.23& 0.35& 0.59& 1.00& 1.83& 5.39& 11.08& 138.65\\
10& 0.54& 0.50& 0.55& 0.60& 0.68& 0.91& 1.47& 2.76& 5.63& 11.57& 174.56\\
12& 0.52& 0.54& 0.57& 0.63& 0.73& 1.00& 1.48& 2.79& 5.84& 11.90& 91.31\\
14& 0.49& 0.49& 0.53& 0.56& 0.69& 0.89& 1.32& 2.62& 5.68& 10.90& 94.71\\
16& 0.41& 0.46& 0.47& 0.56& 0.65& 0.89& 1.29& 2.34& 5.15& 10.74& 19.01\\
\hline
\end{tabular}

%% file: experiments/zdds/table.tex
\begin{tabular}{|r@{~~~}r@{~~~}r@{~~~}r@{~~~}r@{~~~}r@{~~~}r@{~~~}r@{~~~}r@{~~~}r@{~~~}r@{~~~}r@{~~~}r|}
\hline
{\small strategy}&{\small cores}
& {\small ph10}& {\small ph11}& {\small ph12}& {\small ph13}& {\small ph14}& {\small ph15}& {\small ph16}& {\small ph17}& {\small ph18}& {\small ph19}& {\small ph20}\\
\hline
\multirow{3}{*}{{\small original}}&1& 40& 218& 1231& --& --& --& --& --& --& --& --\\
&8& 5& 21& 117& 673& --& --& --& --& --& --& --\\
&16& 3& 13& 64& 357& 2002& --& --& --& --& --& --\\
\hline
\multirow{3}{*}{{\small node}}&1& 2& 3& 6& 10& 15& 26& 43& 64& 99& 147& 209\\
&8& 2& 4& 7& 11& 17& 26& 44& 66& 100& 146& 206\\
&16& 3& 5& 9& 13& 22& 35& 55& 83& 124& 182& 260\\
\hline
\multirow{3}{*}{{\small clause}}&1& 53& 351& 2108& --& --& --& --& --& --& --& --\\
&8& 7& 39& 215& 1305& --& --& --& --& --& --& --\\
&16& 5& 24& 120& 713& --& --& --& --& --& --& --\\
\hline
\end{tabular}